\documentclass[twocolumn,superscriptaddress,nopacs,aps,prl]{revtex4-2}

\usepackage{amssymb}
\usepackage{textgreek}
\usepackage{amsmath}
\usepackage{graphicx}
\usepackage{bm}
\usepackage{color}
\usepackage{xcolor}
\usepackage[normalem]{ulem}
\usepackage{physics}
\usepackage{subfigure}
\usepackage{natbib}
\usepackage{hyperref}
\usepackage{siunitx}

\newcommand{\kB}{k_\mathrm{B}}

\begin{document}

\title{Thermal Effects in the Casimir Torque between Birefringent Plates}

\author{Benjamin Spreng}
\affiliation{Department of Electrical and Computer Engineering, University of California, Davis, California 95616, USA}

\author{Jeremy N. Munday}
\affiliation{Department of Electrical and Computer Engineering, University of California, Davis, California 95616, USA}

\date{\today}

\begin{abstract}
The Casimir effect, originating from quantum and thermal fluctuations, is well known for inducing forces between closely spaced surfaces. When these surfaces are optically anisotropic, these interactions can produce a Casimir torque that rotates the surfaces relative to each other. We investigate, for the first time, the influence of thermal fluctuations on the Casimir torque between birefringent plates. Our results reveal that thermal modes significantly diminish the torque, with reductions up to 2 orders of magnitude for highly birefringent materials. Temperature is also shown to alter the angular dependence of the torque, significantly deviating from the typical sinusoidal behavior, and becomes particularly important at large separations that exceed the thermal wavelength. Finally, we demonstrate that systems of dissimilar birefringent plates that exhibit a distance-dependent reversal in the torque's direction can enable precise control of the torque's magnitude and sign through temperature manipulation. These findings advance our understanding of quantum and thermal fluctuation interplay and provide a framework for designing innovative nanoscale sensors and devices leveraging Casimir torque phenomena.
\end{abstract}

\maketitle

The Casimir effect, arising from zero-point fluctuations of the electromagnetic field, typically manifests as an attractive force between closely spaced macroscopic surfaces~\cite{bordag_advances_2009}. Initially, Casimir predicted the pressure $P$ between two parallel, perfectly conducting planes in vacuum at zero temperature to be $P=-\hbar c \pi^2/(240 d^4)$, where $d$ is the separation between the planes~\cite{Casimir_1948}. This theory has since been extended to account for temperature effects, dispersive materials, and intervening media~\cite{lifshitz_theory_1956, dzyaloshinskii_general_1961}.

At room temperature, thermal corrections to the Casimir force become significant, especially when different models of the dielectric properties of the interacting surfaces are considered. For instance, when modeling gold plates using the Drude model with a finite damping frequency, the thermal correction is repulsive, reducing the Casimir pressure by up to 31\% at micrometer separations~\cite{bostrom_thermal_2000, klimchitskaya_current_2022}. In contrast, if the damping frequency is set to zero, the correction is attractive. Additionally, for a sphere and a plane, a repulsive thermal correction due to geometry has been predicted; however, this effect is less pronounced, reducing the force by up to 3\% when the sphere radius is small compared to the separation~\cite{canaguier-durandThermalCasimirEffect2010a}.

Advances in nanotechnology have enabled high-precision measurements of the Casimir force, facilitating quantitative comparisons with theoretical models~\cite{klimchitskaya_casimir_2009, Decca2011, Lamoreaux2011}. In systems involving graphene, the predicted large thermal corrections are consistent with experimental results~\cite{liuDemonstrationUnusualThermal2021}. However, discrepancies persist for metallic surfaces: most measurements contradict theoretical predictions based on tabulated optical data extrapolated to zero frequency using the dissipative Drude model. Interestingly, the experimental data align well when the dissipationless plasma model is employed instead~\cite{klimchitskaya_recent_2020, klimchitskaya_current_2022}.

For two anisotropic surfaces, fluctuating fields can induce a Casimir torque that tends to rotate the surfaces relative to each other~\cite{kats_van_1971, parsegian_dielectric_1972, barash_moment_1978, mundayTorqueBirefringentPlates2005, mundayErratumTorqueBirefrigent2008}. The Casimir torque has been experimentally verified only once, in an experiment between a birefringent plate and a liquid crystal~\cite{somersMeasurementCasimirTorque2018}. A recent study on the self-alignment of two triangular platelets claims that the Casimir lateral force and torque drive the formation of dimers~\cite{kucukoz_quantum_2024a}. However, quantitative agreement between theoretical predictions and experimental data was not achieved, possibly due to additional electrostatic interactions. Enhancements of the Casimir torque have been predicted using intervening media~\cite{somers_casimirlifshitz_2017} or lamellar gratings~\cite{antezza_giant_2020}, and a distance-dependent sign reversal has been reported~\cite{thiyamDistanceDependentSignReversal2018}, which could be useful for optical observations. For a review of recent developments on the Casimir torque, see Ref.~\cite{sprengRecentDevelopmentsCasimir2022}.

Thermal effects on the Casimir torque have not been previously discussed in the literature. In this Letter, we aim to fill this gap by studying the influence of thermal fluctuations on the Casimir torque between birefringent plates.
We find that thermal modes exert a Casimir torque that counteracts the torque originating from quantum vacuum fluctuations.
This reduction occurs for common birefringent surfaces and can be particularly pronounced for strongly birefringent materials like BaTiO$_3$,
reducing the interaction by up to 99.5\%{}.
Moreover, we observe that the torque between two BaTiO$_3$ plates at an angle $\theta$ no longer displays the typical $\sin(2\theta)$ behavior when thermal fluctuations dominate the interaction.
Additionally, we investigate the high-temperature limit of the Casimir torque, which becomes relevant at separations larger than the thermal wavelength.
We find that strong birefringence is not the only factor contributing to a strong torque; equally important is that the dielectric function perpendicular to the optic axis of the birefringent plate has a moderate value.
Finally, we study a dissimilar system of birefringent plates that undergoes a sign change of the torque with respect to separation. We demonstrate that if the separation is fixed near the transition point where the torque changes sign, the magnitude and sign of the torque can be controlled solely by changing the temperature.

\begin{figure}[t]
\centering
\mbox{\includegraphics[width=\linewidth]{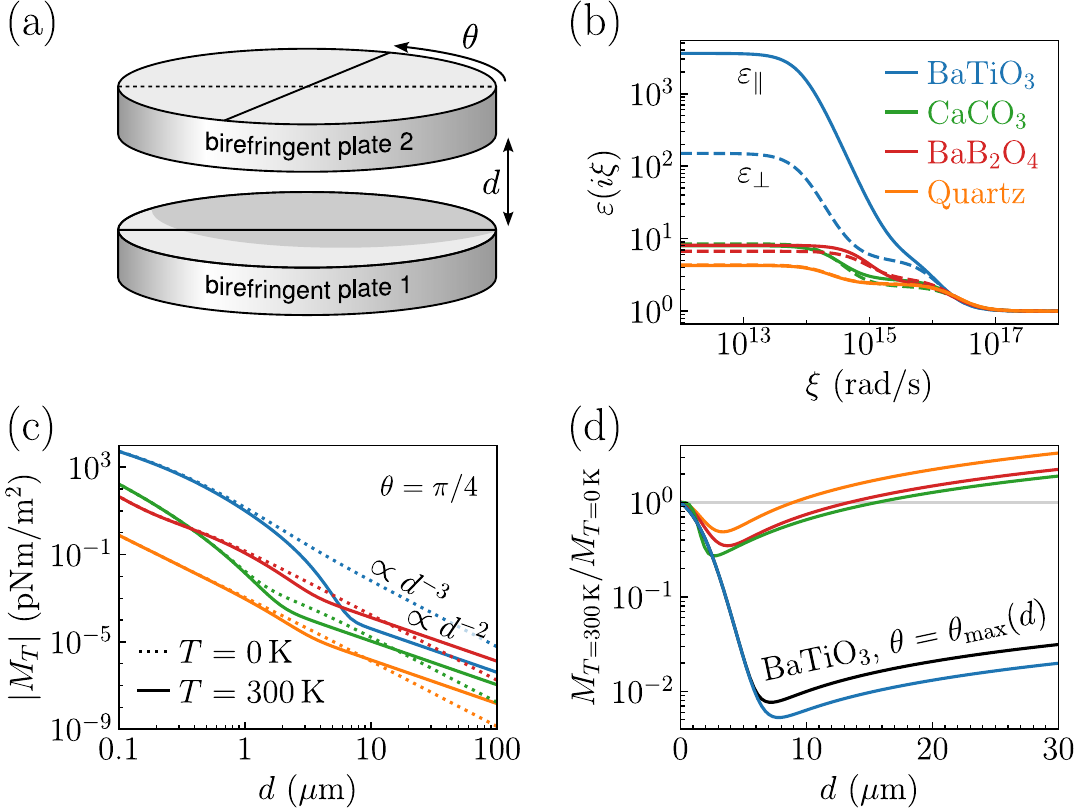}}
\caption{(a) Geometry of two birefringent plates at separation $d$. The optic axes (solid lines) are at a twist angle $\theta$. (b) Dielectric function of BaTiO$_3$, CaCO$_3$, BaB$_2$O$_4$, and quartz as a function of imaginary frequency $\xi$. The component parallel to the optic axis $\varepsilon_\parallel$ is depicted by a solid line, and the perpendicular component $\varepsilon_\perp$ is depicted by a dashed line. (c) Magnitude of the Casimir torque $\vert M_T \vert$ between two plates of the same material as a function of separation $d$ at fixed $\theta=\pi/4$ and fixed temperature of $300\,$K (solid) and $0\,$K (dotted). (d) Ratio of the Casimir torque at $T=300\,$K and $T=0\,$K, respectively, as a function of separation $d$. The black line represents BaTiO$_3$ with the twist angle chosen such that it maximizes the torque (see Fig. 2). For all other curves, the twist angle is set to $\theta=\pi/4$. The gray horizontal line marks the values in which the ratio is unity.}
\label{fig:fig1}
\end{figure}

Consider two birefringent plates for which the optic axes are at a twist angle $\theta$ and the plates' surfaces are separated by a distance $d$ as depicted in Fig.~\ref{fig:fig1}(a).
We choose a coordinate system in which the $z$ axis is perpendicular to the plates' surfaces and the $x$ axis aligns with the optic axis of birefringent plate 1.
Then, the dielectric tensors of birefringent plates 1 and 2 are described by the matrices
\begin{equation}\label{eq:dielectric_tensor}
\begin{pmatrix}
\varepsilon_{1\parallel} & 0 & 0 \\
0 & \varepsilon_{1\perp} & 0 \\
0 & 0 & \varepsilon_{1\perp}
\end{pmatrix}\,,\ 
R_z^\mathsf{T}(\theta)\begin{pmatrix}
\varepsilon_{2\parallel} & 0 & 0 \\
0 & \varepsilon_{2\perp} & 0 \\
0 & 0 & \varepsilon_{2\perp} \end{pmatrix} R_z(\theta)\,,
\end{equation}
respectively. The subscripts $\parallel$ and $\perp$ denote the respective components of the dielectric tensors parallel and perpendicular to the optic axis, and $R_z(\theta)$ is the rotation matrix about the $z$ axis by the angle $\theta$.
The medium between the plates is assumed to be isotropic with dielectric function $\varepsilon_3$. In general, the components of the dielectric tensors \eqref{eq:dielectric_tensor} and $\varepsilon_3$ are dispersive and depend on the angular frequency $\omega$.

With the system being at thermal equilibrium at a temperature $T$, the Casimir torque $M_T$ between the two plates can be expressed  as the negative derivative of the Casimir free energy $\mathcal{F}_T$ with respect to the twist angle $\theta$,
\begin{equation}\label{eq:CasimirFreeEnergy}
M_T = - \frac{d \mathcal{F}_T}{d\theta}\,,\quad\quad \mathcal{F}_T = \kB T \sum_{n=0}^\infty {}' f(\xi_n)
\end{equation}
with the Matsubara frequencies $\xi_n = 2\pi n \kB T/\hbar$ and the prime indicating that the $n=0$ term is weighted with a factor of $1/2$. A negative sign of the torque means that it favors a configuration of decreasing twist angle and vice versa for a positive sign.
In the limit of a vanishing temperature, $T\rightarrow 0$, the Masubara sum becomes dense and can be replaced by an integral. The expression of the free energy then becomes $\mathcal{F}_{T=0} = \hbar \int_0^\infty d\xi\, f(\xi)\,.$
The function $f$ in \eqref{eq:CasimirFreeEnergy} describes the energy contribution for each mode in terms of imaginary frequencies $\xi=-i\omega$ and is given by
\begin{equation}\label{eq:f}
f(\xi) = \frac{1}{4 \pi^2} \int_0^{2\pi}d\varphi \int_0^\infty dk k \log\det \mathbf{D}(\xi, k, \varphi)
\end{equation}
with the $2\times2$ matrix
\begin{equation}\label{eq:D}
\mathbf{D}(\xi, k, \varphi) = \mathbf{1} - \mathbf{r}_1(\varphi) \mathbf{r}_2(\varphi+\theta) e^{-2 \rho_3 d}
\end{equation}
and $\rho_3 = \sqrt{\varepsilon_3 \xi^2/c^2 + k^2}$ and the $2\times2$ reflection matrices $\mathbf{r}_1$ and $\mathbf{r}_2$ for each mode incident on the two birefringent plates, respectively. The entries of the reflection matrices correspond to $p$ and $s$ polarization. When the incidence plane is at an angle with the optic axis of the birefringent plate, polarization may mix upon reflection, and the off-diagonal elements of its reflection matrices are then nonzero. We assume the plates are thick enough to be described by half-spaces. An explicit expression of the reflection matrix for a diagonal dielectric tensor can be found in the Supplemental material. Note that this expression needs to be evaluated at an angle $\varphi+\theta$ for plate 2 to reflect the twist angle between the optic axes.

To study the effect of temperature on the Casimir torque, we first consider systems where the two plates are made of the same birefringent material and the medium between the plates is vacuum ($\varepsilon_3 \equiv 1$). Specifically, we consider the materials BaTiO$_3$, BaB$_2$O$_4$, CaCO$_3$, and quartz. The components of their dielectric tensors are described by Lorentz oscillators with parameters for BaTiO$_3$, CaCO$_3$, and quartz from Ref.~\cite{mundayTorqueBirefringentPlates2005} and parameters for BaB$_2$O$_4$ from Ref.~\cite{somers_conditions_2017}. They are depicted in Fig.~\ref{fig:fig1}(b) as a function of imaginary frequency. Notice that for the sake of studying the effects of thermal fluctuations, we assume the dielectric function is independent of temperature.

Figure~\ref{fig:fig1}(c) depicts the Casimir torque for the four systems as a function of separation $d$ at fixed twist angle $\theta=\pi/4$ for $T=0\,\si{\kelvin}$ (dotted) and $T=300\,\si{\kelvin}$ (solid). For all systems, regardless of temperature, the Casimir torque monotonically decreases with increasing separation. At zero temperature, the BaTiO$_3$ system exhibits the strongest torque at all separations, while quartz shows the weakest torque. At large separations, the zero-temperature torque curves follow a power law proportional to $1/d^3$. This power law can be understood by the fact that the wave-vector components of the confined modes between the plates and thus $\xi$ and $k$ scale as $1/d$ for dimensional reasons. The curves then transition into this power law when the characteristic frequency $\xi_c=c/d$ becomes much smaller than the material's smallest resonance frequency.

The Casimir torque at room temperature agrees with the zero-temperature torque at separations shorter than the thermal wavelength $\lambda_T = \hbar c/(\kB T) \approx 7.6\,\mu$m as expected. With increasing separation, the Casimir torque at room temperature decreases compared to the zero-temperature torque, indicating that the torque due to thermal modes counteracts the torque exerted by quantum vacuum modes. When the separation becomes much larger than the thermal wavelength, the Casimir torque follows a power law proportional to $1/d^2$, which is given by the zero-frequency term in the sum over Matsubara frequencies, given below. Because of the different power laws at large separations, the curves for the room-temperature Casimir torque will cross those for zero temperature, after which the torque contributions from thermal modes will always enhance the torque from quantum modes.

The reduction of the torque due to the thermal modes is quantified in Fig.~\ref{fig:fig1}(d), where the ratio $M_{T=300\,\si{\kelvin}}/M_{T=0\,\si{\kelvin}}$ is plotted as a function of separation. The reduction is the strongest for the BaTiO$_3$ system, with a minimal value of about $0.005$ corresponding to a $200$-fold reduction of the torque exerted by quantum fluctuations. This reduction for the torque is much stronger than the reduction observed for the Casimir force between two metallic plates \cite{bostrom_thermal_2000, klimchitskaya_current_2022} or a sphere and a plane \cite{canaguier-durandThermalCasimirEffect2010a} for which the corresponding ratio is about 0.69 and 0.76, respectively. To put our results here in a different perspective, reducing the temperature thus enhances the torque signal while simultaneously diminishing noise from Brownian motion, favoring low-temperature experiments for precise Casimir torque measurements.

\begin{figure}[t]
\centering
\mbox{\includegraphics[width=\linewidth]{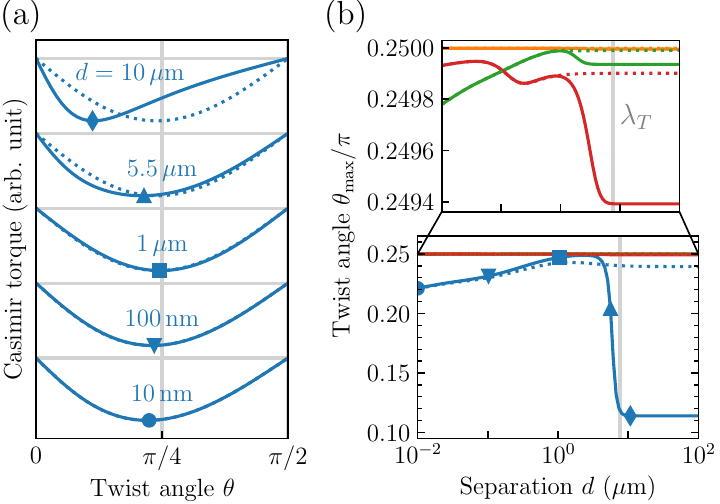}}
\caption{(a) Normalized Casimir torque for two BaTiO$_3$ plates as a function of twist angle $\theta$ at fixed separations as indicated by the labels. The symbols mark the points where the magnitude of the torque is maximal. (b) Twist angle $\theta_\mathrm{max}$ that maximize the magnitude of the Casimir torque as a function of separation $d$. The colors and line style correspond to the same as in Fig.~\ref{fig:fig1}(b). The symbols correspond to the same points as indicated in panel (a). The inset shows a close-up of the vertical axis close to $\theta_\mathrm{max}=\pi/4$. The vertical light gray bar shows the value of the thermal wavelength $\lambda_T=\hbar c/(\kB T)\approx 7.6\,\mu$m at $T=300\,$K.}
\label{fig:fig2}
\end{figure}

Because of the strong birefringence, the Casimir torque between two BaTiO$_3$ plates as a function of twist angle $\theta$ deviates from the $\sin(2\theta)$ behavior known for weakly birefringent plates, as shown in Fig.~\ref{fig:fig2}(a). The deviations are most prominent at larger separations, particularly for the room-temperature torque.
As evident from the upper curves of the torque at $d=10\,\mu$m, the maximum of the torque at room temperature is shifted to a much smaller angle than the one for zero temperature. As a result, the ratio $M_{T=300\,\si{\kelvin}}/M_{T=0\,\si{\kelvin}}$ shown in Fig.~\ref{fig:fig1}(d) at fixed $\theta=\pi/4$ is affected by this shift. For this reason, we include the black curve, which shows the ratio taken for the maximal values of the respective torques. As this curve shows, the torque reduction due to thermal modes is still strong and assumes a minimal value of about $0.008$.

The deviation from the $\sin(2\theta)$ behavior can be characterized by the twist angle $\theta_\mathrm{max}$ at which the magnitude of the torque is maximal [see symbols in Fig.~\ref{fig:fig2}(a)]. In Fig.~\ref{fig:fig2}(b), this twist angle $\theta_\mathrm{max}$ is plotted as a function of separation for the same four systems considered in Fig.~\ref{fig:fig1}. Because of the small birefringence, $\theta_\mathrm{max}$ is close to $\pi/4$ for the systems with BaB$_2$O$_4$, CaCO$_3$, and quartz. Minor deviations can be seen in the inset. Deviations increase at room temperature for large separations $d \gg \lambda_T$ for all systems. At the same time, it is interesting to observe that deviations become stronger at short separations regardless of temperature.

As demonstrated above, thermal effects on the Casimir torque are most prominent at larger separations. In the limit of large separations or equivalently high temperature, $d \gg \lambda_T$, only the zero-frequency term in \eqref{eq:CasimirFreeEnergy} contributes for which only the $p$-polarization-preserving element of the reflection matrices appearing in \eqref{eq:D} is nonvanishing. The integral over $k$ in \eqref{eq:f} for $\xi=0$ can be carried out, and the Casimir free energy then becomes~\cite{somers_conditions_2017}
\begin{equation}
\mathcal{F}_T^{\mathrm{HT}}(d, \theta) = -\frac{\kB T}{32 \pi^2 d^2} \int_0^{2\pi}d\varphi\,\operatorname{Li_3}(r_{1}(\varphi) r_{2}(\varphi+\theta))
\end{equation}
with $\operatorname{Li}_3$ being the polylogarithm of order 3 and
\begin{equation}
r_{j}(\varphi) = \frac{\varepsilon_3 - \varepsilon_{j\perp}\sqrt{1 + (\varepsilon_{j\parallel}/\varepsilon_{j\perp} - 1)\cos^2(\varphi)}}{\varepsilon_3 + \varepsilon_{j\perp}\sqrt{1 + (\varepsilon_{j\parallel}/\varepsilon_{j\perp} - 1)\cos^2(\varphi)}}\,,
\end{equation}
where the dielectric functions are taken in the static limit. The Casimir energy and thus the Casimir torque, $M_T^{\mathrm{HT}} = -\partial \mathcal{F}_T^{\mathrm{HT}}/\partial\theta$, are linear in temperature $T$ and are a $1/d^2$ power law with respect to separation. The dependence on the twist angle $\theta$ and the ratios of dielectric constants $\varepsilon_{j\perp}(0)/\varepsilon_3(0)$ and $\varepsilon_{j\parallel}(0)/\varepsilon_3(0)$ is, in general, more involved.
 
\begin{figure}[t]
\centering
\mbox{\includegraphics[width=\linewidth]{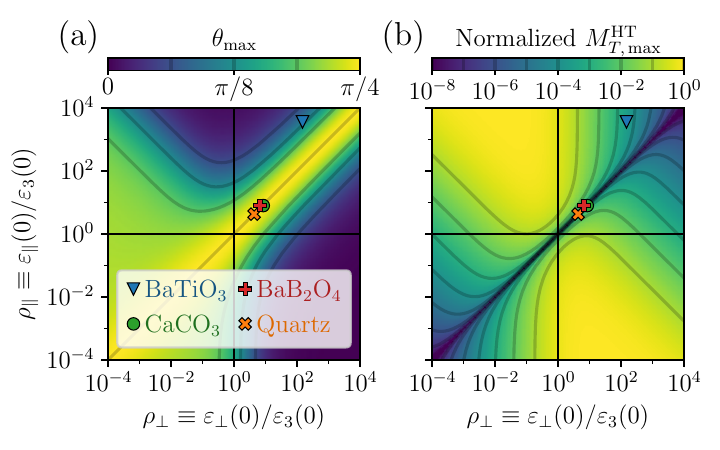}}
\caption{High-temperature limit of the Casimir torque characterized by (a) the twist angle $\theta_\mathrm{max}$ at which the magnitude of the torque is maximal and (b) maximal value of the torque, $M_{T,\mathrm{max}}^\mathrm{HT} \equiv M_{T}^\mathrm{HT}(\theta=\theta_\mathrm{max})$, normalized over the $\varepsilon_\perp(0)/\varepsilon_3(0)$-$\varepsilon_\parallel(0)/\varepsilon_3(0)$ parameter space. The shaded lines indicate contour lines, and the symbols mark the materials studied in Fig.~\ref{fig:fig1}.}
\label{fig:fig3}
\end{figure}

In Fig.~\ref{fig:fig3}, we characterize the Casimir torque in the high-temperature limit $M_T^{\mathrm{HT}}$ for two identical birefringent plates by plotting the twist angle $\theta_\mathrm{max}$ at which the torque is maximal and the normalized value of the maximal torque as a function the ratios of dielectric constants $\rho_\perp\equiv \varepsilon_{\perp}(0)/\varepsilon_3(0)$ and $\rho_\parallel \equiv \varepsilon_{\parallel}(0)/\varepsilon_3(0)$. Ratios smaller than one represent the case where the medium has a dielectric constant larger than the plates. The symbols represent the materials discussed in Fig.~\ref{fig:fig1}. 
A larger parameter space within the figure can be further explored by considering intervening liquids or engineering novel birefringent materials.
The birefringence is zero at the diagonal, and so is the Casimir torque. By defining the anisotropy as the ratio $\Delta = [\varepsilon_{\parallel}(0) - \varepsilon_{\perp}(0)]/[\varepsilon_{\parallel}(0) + \varepsilon_{\perp}(0)]$, the Casimir torque near the diagonal is found to be~\cite{barash_moment_1978}
\begin{equation}\label{eq:low_anisotropy}
M_T^{\mathrm{HT}}(d, \theta) \approx \frac{\kB T}{8\pi d^2} f\left(\frac{\varepsilon_{\perp}(0)}{\varepsilon_3(0)}\right)\Delta^2 \sin(2\theta)
\end{equation}
valid for $\Delta \ll 1$ with
\begin{equation}
f(x) = \frac{x^2}{(1- x^2)^2} \log(1 - \frac{(1-x)^2}{(1 + x)^2})\,.
\end{equation}
Indeed, our numerical results in Fig.~\ref{fig:fig3}(a) show good agreement with \eqref{eq:low_anisotropy} near the diagonal because $\theta_\mathrm{max}$ is close to $\pi/4$. Away from the diagonal, for fixed $\rho_\perp$ ($\rho_\parallel$), $\theta_\mathrm{max}$ decreases with increasing $\rho_\parallel$ ($\rho_\perp$). On the other hand, as depicted in Fig.~\ref{fig:fig3}(b), the value of the maximal torque may stay constant even for extreme values of the dielectric constant ratios, even when $\theta_\mathrm{max}$ goes to zero, as indicated by the contour lines in the right quadrants of the plot, indicating a peculiar torque behavior that is sharply peaked near $\theta=0$ but is otherwise close to zero.

Furthermore, Fig.~\ref{fig:fig3}(b) demonstrates that anisotropy is not the primary factor influencing the strength of the Casimir torque in the high-temperature limit. The ratio $\rho_\perp$ is equally significant. This result is particularly clear in the low-anisotropy expression \eqref{eq:low_anisotropy}, where the maximum torque for each $\Delta$ occurs when the ratio is one, i.e., $\rho_\perp = 1$. For higher levels of anisotropy, the maximum torque shifts to ratio values less than one if the anisotropy $\Delta$ is positive, and greater than one if $\Delta$ is negative. This explains why, as shown in Fig.~\ref{fig:fig1}(c), at room temperature and large separations, the Casimir torque for BaTiO$_3$ [$\Delta=0.92$, $\varepsilon_\perp(0)=150$] is significantly smaller than that for BaB$_2$O$_4$ [$\Delta=0.095$, $\varepsilon_\perp(0)=6.7$], despite BaTiO$_3$ having much larger anisotropy.

\begin{figure}[t]
\centering
\mbox{\includegraphics[width=\linewidth]{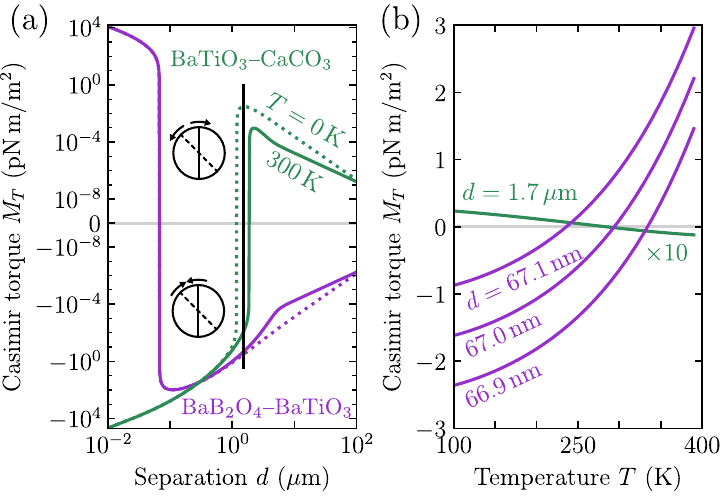}}
\caption{Casimir torque $M_T$ between two dissimilar birefringent plates: BaB$_2$O$_4$-BaTiO$_3$ (purple) and BaTiO$_3$-CaCO$_3$ (green). The twist angle between the optic axes is set to $\theta=\pi/4$. For negative values, the torque acts such that the twist angle decreases and the optic axes align, and vice versa for positive torque values. (a) Casimir torque as a function of separation $d$ at fixed temperature $T=0\,$K (solid) and $T=300\,$K (dotted). (b) Casimir torque as a function of temperature for fixed separation as indicated by the labels. The torque values for the BaTiO$_3$-CaCO$_3$ system are magnified by a factor of 10 to increase visibility.}
\label{fig:fig4}
\end{figure}

Finally, we consider systems of two dissimilar plates. In such systems, the Casimir torque can reverse sign with separation when there is a crossover of the dielectric function components $\varepsilon_\parallel$ and $\varepsilon_\perp$ at a certain imaginary frequency on one birefringent plate but not on the other \cite{thiyamDistanceDependentSignReversal2018}. For the materials considered in this Letter, there is a crossover point for BaB$_2$O$_4$ at $1.51\times 10^{15}\,$rad/s and for BaTiO$_3$ at $2.17\times 10^{16}\,$rad/s, while CaCO$_3$ has two crossover points at $2.69\times 10^{14}\,$rad/s and $2.90\times 10^{16}\,$rad/s. In Fig.~\ref{fig:fig4}(a), we consider the Casimir torque of the two systems BaB$_2$O$_4$-BaTiO$_3$ (purple lines) and BaTiO$_3$-CaCO$_3$ (green lines) in vacuum at fixed twist angle $\pi/4$ as a function of separation $d$. The transition separation at which the torque changes sign depends on temperature. For instance, for the BaTiO$_3$-CaCO$_3$ system, the transition separation at zero temperature is at $1.17\,\mu$m, while it is at $1.87\,\mu$m for room temperature. This allows controlling the sign and magnitude of the Casimir torque at a fixed separation by changing the temperature alone. For example, as shown in Fig.~\ref{fig:fig4}(b) if we fix the separation at $d=1.7\,\mu$m for the BaTiO$_3$-CaCO$_3$ system, we can tune the Casimir torque from $24\,$fNm/m$^2$ at $100\,$K to $-12\,$fNm/m$^2$ at $400\,$K. Similarly, the Casimir torque can be tuned by temperature for the BaB$_2$O$_4$-BaTiO$_3$ system at fixed separation. While the torque values are higher here compared to the other system, the transition separation is more sensitive. The torque values for the BaB$_2$O$_4$-BaTiO$_3$ system are within the range of a proposed Casimir torque experiment \cite{guerout_casimir_2015}.
Our results may be used as a guide to design a separation sensor based on measuring the Casimir torque.

We have studied thermal effects on the Casimir torque between birefringent plates and found that they are more pronounced as compared to thermal effects on the Casimir force between isotropic plates. For instance, the Casimir torque exerted by quantum fluctuations is reduced by thermal fluctuations by 2 orders of magnitude between BaTiO$_3$ plates at room temperature. We have shown that thermal effects also have a significant impact on how the Casimir torque changes as a function of the twist angle between the optic axes of the two plates. Furthermore, we have studied the Casimir torque in the high-temperature limit for a vast parameter space and found that the anisotropy is not the main factor that determines the magnitude of the torque, and the dielectric function component in the direction perpendicular to the optic axes plays a crucial role. Finally, we have demonstrated that in systems that exhibit a sign reversal of the Casimir torque with separation, this reversal may be exploited to change the magnitude and sign of the Casimir torque by changing the temperature of the system alone. Our findings not only contribute to our understanding of the interplay between quantum and thermal fluctuations but also inspire the design of novel sensors with applications to nano- and microelectromechanical devices.

\textit{Acknowledgments}---The authors acknowledge partial support from the National Science Foundation under Grant No. PHY-1806768.

\textit{Data availability}--The data that support the findings of this article are not publicly available. The data are available from the authors upon reasonable request.

\bibliography{references}

\end{document}